# Current driven second harmonic domain wall resonance in ferromagnetic metal/ nonmagnetic metal bilayer: a field free method for spin Hall angle measurements

M. R. Hajiali[1, †], M. Hamdi[2, †], S. E. Roozmeh[1], S. M. Mohseni[2,*]

[1]*Department of Physics, University of Kashan, 87317 Kashan, Iran*
[2]*Faculty of Physics, Shahid Beheshti University, Evin, 19839 Tehran, Iran*

We study the ac current-driven domain wall motion in bilayer ferromagnetic metal (FM)/nonmagnetic metal (NM) nanowire. The solution of the modified Landau-Lifshitz-Gilbert equation including all the spin transfer torques is used to describe motion of the domain wall in presence of the spin Hall effect. We show that the domain wall center has second harmonic frequency response in addition to the known first harmonic excitation. In contrast to the experimentally observed second harmonic response in harmonic Hall measurements of spin-orbit torque in magnetic thin films, this second harmonic response directly originates from spin-orbit torque driven domain wall dynamics. Based on the spin current generated by domain wall dynamics, the longitudinal spin motive force generated voltage across the length of the nanowire is determined. The second harmonic response introduces additionally a new practical field-free and all-electrical method to probe the effective spin Hall angle for FM/NM bilayer structures that could be applied in experiments. Our results also demonstrate the capability of utilizing FM/NM bilayer structure in domain wall based spin torque signal generators and resonators.

## I. INTRODUCTION

Current induced domain wall (DW) dynamics in magnetic nanostructures raised great research interest in the field of spintronics because of its perspective applications in novel devices including DW based shift register [1], logic [2], racetrack memory [3] and DW collision spin wave emitter [4]. Current-induced DW motion (CIDWM) in magnetic nanowires can occur thanks to the spin transfer torque (STT) effect due to exchange coupling between local magnetization of DW and spin-polarized currents [5–9]. There are two types of STTs acting on a non-collinear magnetization texture, e.g. a DW, when a spin-polarized current flows through it, namely adiabatic and non-adiabatic terms. It has been shown theoretically [5,10] that the initial DW velocity is mostly controlled by the adiabatic term, while the non-adiabatic term is responsible for its terminal velocity.

In ferromagnetic metal/nonmagnetic metal (FM/NM) bilayer, a spin current generated by the spin Hall effect (SHE) through the adjacent NM layer can be injected into the FM layer [11] that produces another type of STT, named spin-orbit torque (SOT) which can result in magnetization dynamics and CIDWM [12–16]. For instance, SOT can significantly reduce the critical dc current density to depinning a DW from a pinning potential, therefore causes lowering of the energy consumption for operation of DW-based electronic devices [17].

Investigation of dynamical response of DW to ac currents have resulted in many achievements including determination of the mass of DW [18,19], micrometer range DW displacement [18,19] and resonant control of DW movement [20–22]. Furthermore, depinning of a DW benefits more from ac current than the dc one, because of current induced DW resonance [23]. Therefore, to benefit of such dynamically rich feature, uncovering dynamics of DW under ac current and SOT is important and remained unclear.

Since the discovery of SHE in semiconductors [24,25] and metals [26,27], various techniques have been developed to determine the conversion rate of charge currents to spin currents named spin Hall angle, $\theta_{SH}$. All these methods including cavity ferromagnetic resonance (C-FMR) spectrometer [28,29], spin pumping (SP) [30], spin torque FMR (ST-FMR) [31,32], hybrid phase-resolved optical-electrical FMR (OE-FMR) [33] and determining the DW velocity using magneto-optical microscopy [34,35] require almost strong bias magnetic fields to saturate the FM layer or in some cases a complicated measurement setup. Hence, a field-free measurement technique of $\theta_{SH}$ is a challenge.

In this paper, we investigate the ac CIDWM in FM/NM bilayer (Fig.1) based on collective coordinate approach. FM layer has perpendicular magnetic anisotropy and NM has strong spin-orbit interaction responsible for the SHE. All current induced STTs and the SOT are included in our study. By solving equations of DW motion, we obtained second harmonic motion in addition to the known first harmonic motion. This second harmonic response has different nature in comparison with observed one in second harmonic Hall measurements of SOT [36–41] in single domain FM/NM structures. The former that we study here originates from SOT exerted on DW due to ac SHE occurred in NM layer, while the latter originates from anomalous Hall and spin Hall magnetoresistance effects [36–41].

Finally, we calculated the spin motive force (SMF) voltage induced by DW motion. Our results propose a new field-free method to measure effective spin Hall angle for such bilayer structures. In addition, second harmonic DW motion in presence of SHE and ac current can be used further in recently reported DW and skyrmion based high frequency signal generators and resonators [20,21,42] to obtain higher frequencies.

## II. MODEL
### A. ac current induced domain wall motion in presence of spin Hall effect

We considered a bilayer strip wire with dimensions $2L \times w \times (t_F + t_N)$ along x, y and z directions, where $2L$ is the length and $w$ is the width of bilayer wire, $t_F$ and $t_N$ represent the thickness of FM and NM layers, respectively and $L \gg w, t_F, t_N$. An in-plane ac current applied along the x-direction generates conventional adiabatic and non-adiabatic STTs and SOT acting on a DW within FM. The geometry of the structure is shown in Fig. (1). The modified Landau-Lifshitz-Gilbert (LLG) equation including all the STTs is given by

$$\frac{\partial \mathbf{m}}{\partial t} = -\gamma \mathbf{m} \times \mathbf{H}_{eff} + \alpha \mathbf{m} \times \frac{\partial \mathbf{m}}{\partial t} - a_J \mathbf{m} \times \left( \mathbf{m} \times \frac{\partial \mathbf{m}}{\partial t} \right) - n_J \mathbf{m} \times \frac{\partial \mathbf{m}}{\partial x} - \theta_{SH} c_J \mathbf{m} \times (\mathbf{m} \times \hat{y}), \quad (1)$$

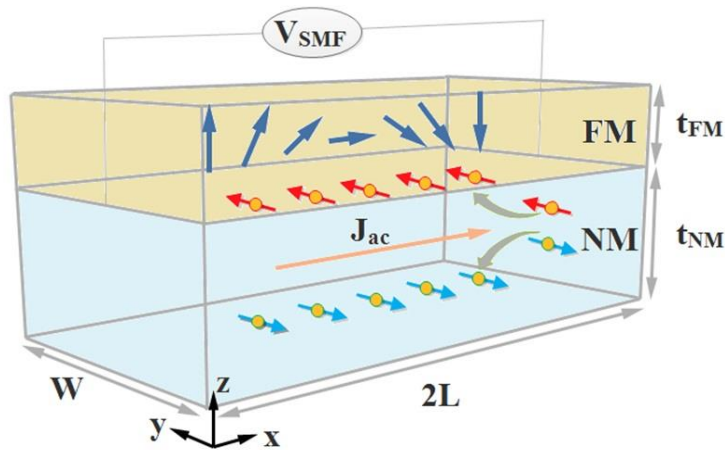

FIG. 1. Schematic illustration of a FM/NM metal bilayer system. An in-plane ac current density $J_{ac}$ generates a perpendicular spin current, which exerts SOT on FM.

where, **m** is the unit vector along the magnetization direction, $\gamma$ is the gyromagnetic ratio, **H**$_{eff}$ is effective field including the exchange, anisotropy and external fields, $\alpha$ is the Gilbert damping constant, $a_J = (\hbar\gamma P/2eM_s)J_F$ is magnitude of adiabatic STT in velocity dimension, $n_J = \beta a_J$ is magnitude of non-adiabatic STT and $\beta$ is the non-adiabaticity coefficient, $\theta_{SH}c_J = (\theta_{SH}\hbar\gamma J_N/2eM_s t_F)$ is magnitude of SOT, where $\theta_{SH}$ is an effective spin Hall angle for bilayer system, $\hbar$ is the reduced Planck constant, $P$ is the spin polarization in FM, $e$ is electron charge, $M_s$ is saturation magnetization of FM, and $J_F(J_N)$ is applied current density in FM (NM). $J_F$ and $J_N$ are determined by Ohms-law for two parallel resistors as $J_F = J(\frac{\sigma_F}{\sigma_F+\sigma_N})$ and $J_N = J(\frac{\sigma_N}{\sigma_F+\sigma_N})$, where $J$ is average current density in bilayer nanowire and $\sigma_F(\sigma_N)$ is conductivity of FM (NM).

Introducing the magnetization direction as $\mathbf{m} = (\sin\theta \cos\varphi, \sin\theta \sin\varphi, \cos\theta)$, the micromagnetic energy density $u$ in polar coordinates $\theta$ and $\varphi$ is given by

$$u = A_{ex}\left[\left(\frac{\partial\theta}{\partial x}\right)^2 + \left(\sin\theta\frac{\partial\varphi}{\partial x}\right)^2\right] + K_u \sin^2\theta + K_d \sin^2\theta \sin^2\varphi, \quad (2)$$

where $A_{ex}$ is exchange stiffness coefficient and $K_u$ and $K_d$ are perpendicular and in-plane anisotropy constants, respectively. Dynamic DW structure is considered as $\theta(x,t) = 2\arctan\left[\exp\left(\frac{x-q(t)}{\Delta(t)}\right)\right]$ and $\varphi(x,t) = \phi(t)$, where $q(t)$ and $\Delta(t)$ represent DW center position and width, respectively [43]. Chirality of the DW is determined by angle $\phi(t)$ and distortion of wall width is small during wall motion, hence we suppose $\Delta(t) \cong \sqrt{A/K_u}$. The restoring energy density arising from geometric notch [44, 45], impurity or demagnetization field [23] induced pinning is assumed to be of the form $u_r = k\,q^2/2$. Therefore, equations of motion for two collective coordinates $q$ and $\phi$ in rigid DW limit, by neglecting the small coefficients of nonlinear terms, are given by

$$m\frac{\partial^2 q}{\partial t^2} + b\frac{\partial q}{\partial t} + kq = -\frac{2M_s}{\gamma}\left(\frac{n_J}{\Delta} + \frac{1}{\gamma\Delta H_d}\left(\alpha\frac{\partial n_J}{\partial t} + \frac{\partial a_J}{\partial t}\right) + \frac{\alpha a_J n_J B_{SH}}{\gamma\Delta^2 H_d}\right), \quad (3)$$

$$m\frac{\partial^2\phi}{\partial t^2} + b\frac{\partial\phi}{\partial t} + k\phi = \frac{ka_J}{\gamma\Delta H_d} + \frac{2M_s\alpha}{\Delta^2\gamma^2 H_d}\frac{\partial a_J}{\partial t} - \frac{2M_s}{\Delta^2\gamma^2 H_d}\frac{\partial n_J}{\partial t}, \quad (4)$$

where $H_d = (2K_d/M_s) = 4\pi M_s$ is demagnetization field, $B_{SH} = \theta_{SH}c_J(\pi\Delta eM_s/PJ_F\hbar\gamma) = (\pi\theta_{SH}\Delta J_N/2t_F PJ_F)$, $m = (1+\alpha^2)\frac{2M_s}{\gamma^2\Delta H_d}$ and $b = \alpha(\frac{2M_s}{\gamma\Delta} + \frac{k}{\gamma H_d})$.

Considering an ac applied current with frequency ω, Eq. (3) describes forced damped oscillations. Therefore, the steady state solution to $q$ is of the form

$$\begin{aligned} q(t) &= q_{1\omega}(t) + q_{2\omega}(t) \\ q_{1\omega}(t) &= A_\omega \cos[\omega t - (\delta - \rho)] \\ q_{2\omega}(t) &= A_{2\omega} \cos[2\omega t - \xi] \end{aligned} \quad (5)$$

where $q_{1\omega}(t)$, $q_{2\omega}(t)$ are the first and second harmonic components of DW motion. $A_\omega$ and $A_{2\omega}$ are the first and second harmonic DW motion amplitudes, respectively, and are given by

$$A_\omega = -\frac{2M_S}{\gamma}\left(\frac{\sqrt{\frac{n^2}{\Delta^2}+\frac{(\alpha n+a)^2 \omega^2}{(4\pi\gamma\Delta M_S)^2}}\left(\frac{\sigma_F}{\sigma_F+\sigma_N}\right)J}{\sqrt{(k-m\omega^2)^2+\omega^2 b^2}}\right) \quad (6)$$

$$A_{2\omega} = -\frac{2M_S}{\gamma}\left(\frac{\frac{\alpha n a B_{SH}}{4\pi\gamma\Delta^2 M_S}\left(\frac{\sigma_F}{\sigma_F+\sigma_N}\right)^2 J^2}{\sqrt{(k-4m\omega^2)^2+4\omega^2 b^2}}\right) \quad (7)$$

$\phi(t)$ has no second harmonic dynamics and obtain as

$$\phi(t) = \phi_\omega \cos\left[\omega t - (\delta+\eta)\right] \quad (8)$$

where the chirality amplitude is

$$\phi_\omega = \frac{2M_S}{\gamma}\left(\frac{\sqrt{\left(\frac{ak}{8\pi\Delta M_S^2}\right)^2+\frac{(n-\alpha a)^2 \omega^2}{(4\pi\gamma\Delta^2 M_S)^2}}\left(\frac{\sigma_F}{\sigma_F+\sigma_N}\right)J}{\sqrt{(k-m\omega^2)^2+\omega^2 b^2}}\right) \quad (9)$$

The phases δ, ρ, ξ and η are in the form of

$$\begin{aligned}\delta &= \arctan\left[b\omega/(k-m\omega^2)\right], \\ \rho &= \arctan\left[(\alpha n+a)\omega/(4\pi M_S\gamma n)\right], \\ \xi &= \arctan\left[2b\omega/(k-4m\omega^2)\right], \\ \eta &= \arctan\left[2M_S\omega(n-\alpha a)/\Delta\gamma ak\right].\end{aligned} \quad (10)$$

According to Eqs. (6, 7 and 9) the DW motion amplitude $A_\omega$ and chirality amplitude $\phi_\omega$ are proportional to the applied current density $J$, while the second harmonic amplitude is proportional to the $J^2$ due to ac SOT.

### B. Spin currents and voltages

A given dynamical non-collinear magnetization texture, **m**(r, t), generates a spin current which is given by [46–49]

$$J_i^s = \frac{\mu_B \hbar}{2e^2} \sum_k \sigma_{ik}^c \left[ \left( \frac{\partial \mathbf{m}}{\partial t} \times \frac{\partial \mathbf{m}}{\partial r_k} \right) \cdot \mathbf{m} + \beta \frac{\partial \mathbf{m}}{\partial t} \cdot \frac{\partial \mathbf{m}}{\partial r_k} \right] \quad (11)$$

where $J_i^s$ is the *i'th* component of spin current, $\mathbf{J}^s = (J_x^s, J_y^s, J_z^s)$, which flows through *i*-direction (*i, j, k=x, y, z*) and its polarization is determined by instantaneous local magnetization direction, $\mathbf{m}(r, t)$, $\sigma_{ik}^c$ is the *ik*-element of electrical conductivity tensor of FM, $r_k = x, y, z$ and $\mu_B$ is the Bohr magneton. All other parameters are determined before. Such spin current generates electrical currents or voltages in FM/NM bilayer through two dominant mechanisms: i) spin motive force (SMF) within the FM layer and ii) inverse spin Hall effect (ISHE) within the NM layer. SMF is related to the s-d exchange coupling between conduction electrons and localized moments within FM. While, ISHE is the reciprocal effect of SHE, i.e. an injected spin current into NM is converted to a transverse electrical current or voltage. In the case of our study and considering dynamic DW structure introduced before, the only non-vanishing spatial derivative of magnetization is $\partial \mathbf{m}(r,t)/\partial x$. Furthermore, most of common FM metals have cubic or hexagonal crystal structure and hence their conductivity tensor is diagonal. Therefore, according to Eq. (9), all components of spin current vanish except $J_x^s$ and $\mathbf{J}^s = (J_x^s, 0, 0)$. As the z-direction flowing component of the spin current, $J_z^s = 0$, no spin current enters to the NM layer hence no charge current or voltage is generated through ISHE. The local current density generated by SMF mechanism is given by

$$J_{i,SMF}^c = \frac{eP}{\mu_B} J_i^s, \quad (12)$$

Here, $J_{i,SMF}^c$ is the charge current density flowing through *i*-direction generated by SMF. Using Eq. (9) and Eq. (10) and dynamic DW structure with $\partial \theta / \partial t = -(\partial \theta / \partial x)(\partial q / \partial t)$ and $\partial \theta / \partial x = (\sin \theta / \Delta)$, one can obtain the charge current density induced by SMF in polar coordinates as

$$\begin{aligned} J_{x,SMF}^c &= \frac{P\hbar \sigma_F}{2e} \left[ -\frac{\partial \theta}{\partial x} \frac{\partial \varphi}{\partial t} \sin \theta + \beta \frac{\partial \theta}{\partial x} \frac{\partial \theta}{\partial t} \right] \\ &= -\frac{P\hbar \sigma_F}{2e} \left[ \frac{1}{\Delta} \frac{\partial \phi}{\partial t} + \frac{\beta}{\Delta^2} \frac{\partial q}{\partial t} \right] \sin^2 \theta. \end{aligned} \quad (13)$$

Averaging SMF charge current density over the wire length, $\langle J_{x,SMF}^c \rangle = \frac{1}{2L} \int_{-L}^{L} J_{x,SMF}^c \, dx$ and considering first and second harmonic components of DW motion in Eq. (5), SMF voltage can be decomposed to first harmonic, $V_{SMF}^\omega$, and second harmonic, $V_{SMF}^{2\omega}$, components as

$$\begin{aligned} V_{SMF}^\omega (t) &= V^\omega \sin[\omega t - \psi], \\ V_{SMF}^{2\omega} (t) &= V^{2\omega} \sin[2\omega t - \xi]. \end{aligned} \quad (14)$$

where the voltage amplitudes $V^\omega$ and $V^{2\omega}$ and first harmonic voltage phase, ψ are given by

$$V^{\omega} = \frac{\hbar P\omega}{e\Delta}\sqrt{\beta^2 A_\omega^2 + \Delta^2\phi_\omega^2 + 2\beta\Delta A_\omega\phi_\omega \cos[\rho+\eta]},$$

$$\tan\psi = \frac{\beta A_\omega \sin[\delta-\rho] - \Delta\phi_\omega \sin[\delta+\eta]}{\beta A_\omega \cos[\delta-\rho] + \Delta\phi_\omega \cos[\delta+\eta]},\quad (15)$$

$$V^{2\omega} = \frac{2\hbar P\omega\beta}{e\Delta} A_{2\omega}. \quad (16)$$

### III. RESULTS AND DISCUSSIONS
#### A. DW dynamics

Implementing the realistic geometric and material parameters introduced in Table I and current density value of $J = 20\times10^{13}$ A/m$^2$ [13,15,23,50] into Eq. (5-7), we plot DW motion amplitudes and components. Fig. (2a) shows $A_\omega$ (blue line) and $A_{2\omega}$ (red line) as a function of applied current frequency, ω. There is a resonance peak in both first and second harmonic amplitudes of 390 nm and 30 nm, respectively. Inset shows the second harmonic amplitude separately for clarification. The resonance frequencies for first and second harmonic motion are current independent and for used material parameters they are obtained as $\omega_{r,1}$=59.9 GHz and $\omega_{r,2}$=29.4 GHz, respectively. It is obvious that the second harmonic component of DW motion and its related effects are significant at second harmonic resonance frequency, $\omega_{r,2}$. Therefore, we used $\omega=\omega_{r,2}$ to perform all other calculations hereafter. Fig. (2b) shows the time dependence of first (blue line) and second (red line) harmonic components of and the total (black line) DW motion, respectively.

TABLE I. Realistic material parameters and geometry dimensions for numerical calculations [13,15,23,50].

|  | Symbol | Value | Unit |
|---|---|---|---|
| Saturation magnetization of FM | $M_s$ | 0.5 | T |
| Exchange stiffness coefficient | $A_{ex}$ | 3.37×10$^{-11}$ | J/m |
| Perpendicular anisotropy constants | $K_u$ | 1.5×10$^5$ | J/m$^3$ |
| Gilbert damping constant | $\alpha$ | 0.02 |  |
| Nonadiabaticity of STT | $\beta$ | 0.05 |  |
| Restoring force constant | $k$ | 1.25×10$^6$ | T$^2$/m |
| Conductivity of FM (NM) | $\sigma_F$ | 1.5×10$^7$ | s/m |
|  | ($\sigma_N$) | (1.89×10$^7$) |  |
| Spin polarization in the FM | $P$ | 0.5 |  |
| Spin Hall angle | $\theta_{SH}$ | 0.15 |  |
| Domain wall width | $\Delta$ | 15 | nm |
| Thickness of FM (NM) | $t_F(t_N)$ | 0.6 (2) | nm |
| Applied Current density | $J$ | (5-20)×10$^{13}$ | A/m$^2$ |
| Width of bilayer wire | $w$ | 5 | nm |

| Half-length of bilayer wire | $L$ | $10\times10^{-6}$ | m |

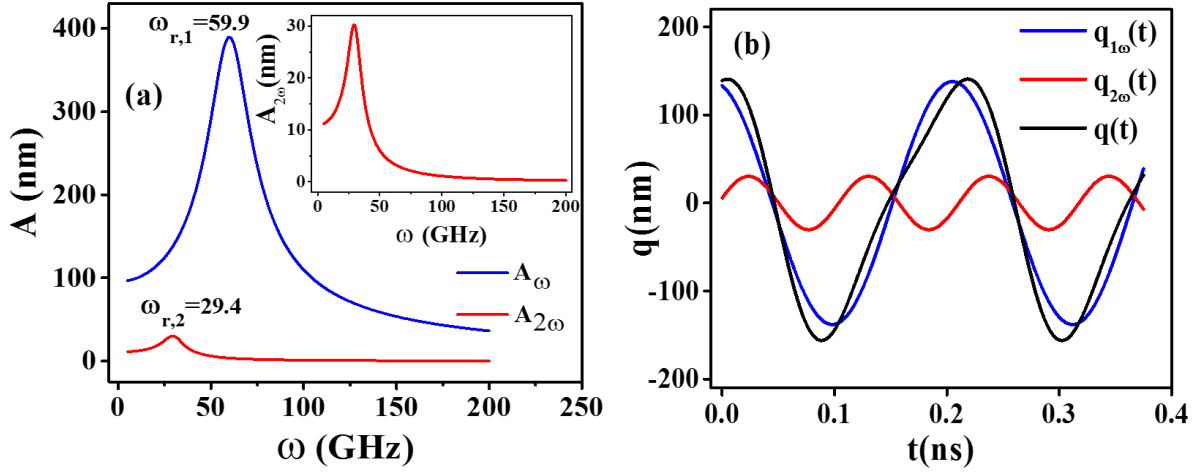

FIG. 2. (a) Frequency dependence of first harmonic ($A_\omega$) and second harmonic ($A_{2\omega}$) amplitudes, respectively. The inset shows the second harmonic motion amplitude separately. (b) Time dependence of first harmonic ($q_{1\omega}$), second harmonic ($q_{2\omega}$) and total ($q$) DW motion at second harmonic resonance frequency, respectively. Applied current density is $20\times10^{13}$ A/m$^2$.

## B. SMF voltages

Using Eq. (15, 16) and material parameters in Table I we obtain the current dependence of first and second harmonic SMF voltage amplitudes at $\omega_{r,2}$ which are shown in Fig. (3a, 3b), respectively. As mentioned before, $A_\omega$ and $\phi_\omega$ linearly depend on applied current, $J$, hence the first harmonic voltage amplitude, $V^\omega$, also depends linearly on $J$. While, second harmonic voltage amplitude, $V^{2\omega}$, quadratically grows with $J$ due to quadratic dependence of $A_{2\omega}$ on $J$.

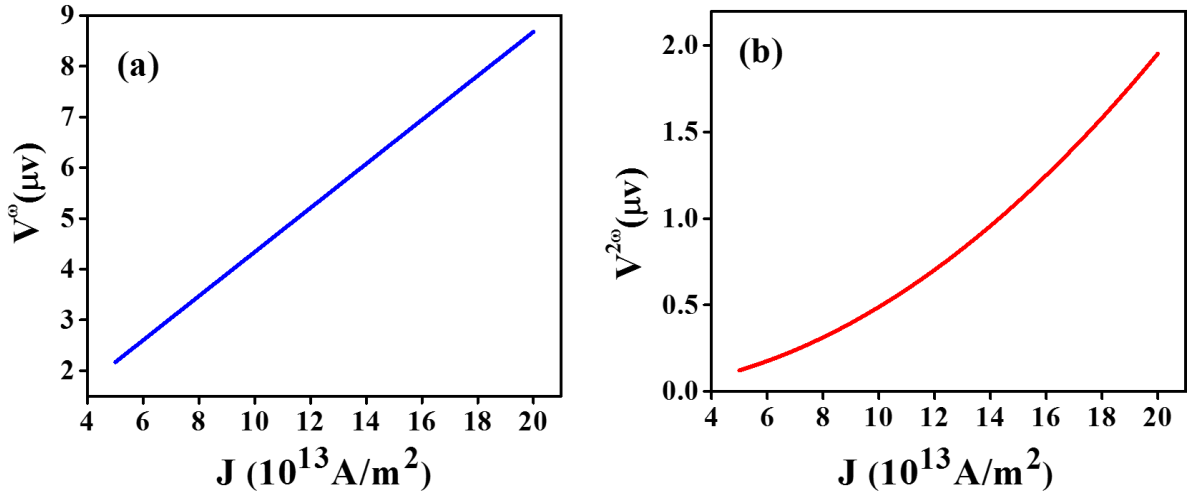

FIG. 3. Applied current density dependence of (a) first harmonic ($V^\omega$) and (b) second harmonic ($V^{2\omega}$) SMF voltage amplitudes at second harmonic resonance frequency.

## C. Field-free spin Hall angle measurement based on electrical detection of second harmonic DW resonance

Based on our results, we propose a new field-free practical experimental method to measure the effective spin Hall angle for a bilayer FM/NM structure based on electrical detection of the current driven second harmonic DW resonance. In this technique, one can apply an ac current with constant amplitude to a bilayer strip of FM/NM structure and sweep the frequency and simultaneously probe the SMF second harmonic voltage along the wire to find second harmonic resonance frequency, $\omega_{r,2}$. Then, by fixing the frequency to second harmonic resonance frequency and sweeping the current amplitude, one thus measures the current dependent second harmonic voltages. Fitting the result with Eq. (16), and using appropriate material parameters, effective spin Hall angle, $\theta_{SH}$, can be determined. This could be a powerful field-free and all-electrical method to measure spin Hall angle for FM/NM structures in comparison with any existing field required techniques such as cavity ferromagnetic resonance (C-FMR) spectrometer [28,29], spin pumping (SP) [30], spin torque FMR (ST-FMR) [31,32], hybrid phase-resolved optical-electrical FMR (OE-FMR) [33]. All these FMR-based techniques require large magnetic fields of the order of few Teslas to saturate the FM layer and provide the FMR conditions. Along with this, the DW velocity measurement based techniques requires a complicated measurement setup such as magneto-optical Kerr microscopy [34,35]. Therefore, our proposed technique represents advantage of both field-free and simple all-electrical measurements. In addition, second harmonic generation in FM/NM structures, by a DW or a non-collinear magnetization texture in general, could be useful in designing DW-based spin torque signal generators and resonators.

## IV. CONCLUSIONS

Using the real material parameters and reasonable current densities, we predict a second harmonic DW resonance in FM/NM bilayer nanowires originated from ac SOT exerted on FM layer. The longitudinal SMF induced second harmonic voltage is calculated in the range of 0.15-2 µV which is measurable in laboratories. In addition, we introduced a new field-free all-electrical method to obtain effective spin Hall angle for FM/NM bilayer structures based on current induced second harmonic DW resonance. Furthermore, our results can help to efficiently designing DW based spin-torque signal generators and resonators.

## ACKNOWLEDGMENTS

We acknowledge support from the Iran Science Elites Federation (ISEF).

† These authors contributed equally.
*Corresponding author: m-mohseni@sbu.ac.ir, majidmohseni@gmail.com.